\pdfoutput=1
\documentclass[epj,twocolumn]{webofc}
\usepackage[varg]{txfonts}   
\newcommand{\vev}[1]{\langle {#1} \rangle}
\woctitle{LHCP 2013}
\begin{document}
\title{A higgs-like dilaton: viability and implications}
%
%

\author{Javi Serra\inst{1}\fnsep\thanks{\email{js993@cornell.edu}} 
}

\institute{Department of Physics, LEPP, Cornell University, Ithaca, NY 14853, USA 
          }

\abstract{%
We re-examine the feasibility of the higgs-like particle discovered at the LHC being a dilaton: the Goldstone boson of spontaneous breaking of scale invariance.
We review the expected phenomenological deviations from the SM higgs and compare with other Goldstone higgs scenarios, with particular emphasis on double higgs production.
}
\maketitle
\section{Introduction}
\label{intro}

The discovery of the long predicted higgs boson has brought unprecedented opportunities to unravel the nature of electroweak (EW) symmetry breaking.
For the time being we rely on the measurement of the new particle's mass and linear couplings to Standard Model (SM) fields.
With this information, it is mandatory to re-evaluate our expectations for the physics giving rise to the EW scale, and indeed already a great deal of repercussions have fallen upon, for instance, minimal supersymmetric models.
Likewise for models of strong dynamics, for which such a low mass, $\sim \! 125 \, \mathrm{GeV}$, is an indication that the scalar must be a composite Goldstone boson of some sort.
Two distinct candidates stand, either the one arising from the spontaneous breaking of an internal global symmetry, which we refer to as the composite higgs
(taking \cite{Agashe:2004rs} as reference), 
or the one from the spontaneous breaking of scale invariance (SBSI), the dilaton.
Here we focus on the latter, following the analysis of \cite{Bellazzini:2012vz}.

The naturalness and hierarchy problems of the EW scale are the guiding principles to postulate a new strongly interacting sector at $\Lambda_{IR} \sim \mathrm{TeV}$.
As we learned from QCD, strong dynamics is able to naturally generate a large hierarchy between two distinct physical scales $\Lambda_{IR} \ll \Lambda_{UV}$, while at the same time avoiding a large sensitivity of the former to the latter.
This is achieved by building a lagrangian with no relevant operators (unless protected by symmetry), but only operators close to marginality or irrelevant.
This is tantamount to an approximate scale invariant regime between the two scales, 
where scale transformations (we denote by $d$ the scaling dimension of a field/operator)
\begin{equation}
x \to e^\alpha x \ , \quad \phi(x) \to e^{d_\phi \alpha} \phi(e^\alpha x) \ ,
\label{scaletrans}
\end{equation}
leave invariant the action,
\begin{equation}
\mathcal{S} = \int \! d^4 x \, \mathcal{L} \ , \quad \mathcal{L} = \sum_{\mathcal{O}} g_{\mathcal{O}} \, \mathcal{O}(\phi,\partial_\mu \phi) \ .
\label{action}
\end{equation}
The lagrangian is written as a sum over operators with $d_\mathcal{O} = 4$,
while the coupling $g_{\mathcal{O}}$ becomes irrelevant in the infrared for those operators with $d_\mathcal{O} \gg 4$.

It is further assumed that the lagrangian contains the proper dynamics for SBSI, in the form of the vacuum condensate of a scalar field, $\vev{\phi} = f^{d_\phi}$, with $\Lambda_{IR} \sim 4 \pi f$.\footnote{It is understood that part or all of the composite fields that obtain a VEV carry EW quantum numbers.}
Naively, such a breaking gives rise to one massless Goldstone boson, the dilaton,
parametrized as $\chi \equiv f e^{\sigma(x)/f}$ ($\chi \to e^\alpha \chi$ under scaling).
However, a more careful treatment of SBSI must take into account that a non-trivial potential for $\chi$ is allowed by the symmetry,
\begin{equation}
V(\chi) = F_0 \, \chi^4 \ ,
\label{dilpot0}
\end{equation}
which makes a natural realization of SBSI highly non-trivial.
As Fubini showed \cite{Fubini:1976jm}, and we summarize in Table~\ref{Fubini}, the only way to obtain a Poincar\'e-4 invariant vacuum given Eq.~(\ref{dilpot0}) is to tune $F_0 = 0$, in which case $\vev{\chi} = f$ remains undetermined (a flat direction).
\begin{table}[h]
\centering
\caption{Patterns of SBSI \cite{Fubini:1976jm}. $r$ is a spatial coordinate while $t$ is time-like.}
\label{Fubini}
\begin{tabular}{ccc}
\hline
$F_0 > 0$ & $F_0 = 0$ & $F_0 < 0$ \\ \hline
AdS-4 & Poincar\'e-4 & dS-4 \\ 
$\vev{\chi} \propto 1/r$ & $\vev{\chi} = f$ & $\vev{\chi} \propto 1/t$ \\
\hline
\end{tabular}
\end{table}

\noindent This is indeed a tuning since naive dimensional analysis yields $F_0 \sim (4 \pi)^2$ (unless supersymmetry is involved).
The problem is even more evident if we notice that $F_0$ determines the size of the effective cosmological constant associated to the SBSI, $V(\vev{\chi}) = F_0 f^4$.
The only way out of this conclusion is the introduction of a perturbation that explicit breaks scale invariance,
since this allows for a stabilization mechanism that naturally produces $V(\vev{\chi}) \simeq 0$.
However, this essential breaking of the symmetry automatically makes the presence of a light (pseudo-)Goldstone from SBSI quite more involved than that of Goldstone's from global symmetries.

The explicit breaking of scale invariance is introduced by the (generic) perturbation
\begin{equation}
\delta \mathcal{L} = \lambda \, \mathcal{O}_\lambda \ , \quad \gamma_{\lambda} \equiv 4-d_{\mathcal{O}_\lambda} = \frac{d \log \lambda}{d \log \mu} = \frac{\beta(\lambda)}{\lambda} \ ,
\label{breaking}
\end{equation}
where the operator's anomalous dimension $\gamma_{\lambda}$ is defined at the quantum level, that is including loop contributions to the $\beta$-function.
As Coleman and Weinberg showed \cite{Coleman:1973jx}, and as it can be derived from a spurious analysis based on scale invariance, the new effective dilaton potential reads
\begin{equation}
V(\chi) = F[\lambda(\chi)] \, \chi^4 \ , 
\label{dilpot}
\end{equation}
where a $\chi$-dependent quartic coupling $F$ has been generated through the scale dependence of the perturbation $\lambda$.
In order for the minimum of (\ref{dilpot}) to be hierarchical, the perturbation must have a small $\beta$-function. 
Even more important, in order for SBSI to be genuine, the $\beta$-function must remain small at the minimum.
Otherwise, the dilaton and all its properties will be absent in the low energy lagrangian.\\

Given this picture, our goal is to understand if the dilaton could really mimic the 125 GeV higgs, Sec.~\ref{sec-1} and \ref{sec-2}, and if so, what the differences with other higgs-like states would be, Sec~\ref{sec-3}. 

\section{Dilaton mass}
\label{sec-1}

\begin{figure}
\centering
\includegraphics[width=4.5cm,clip]{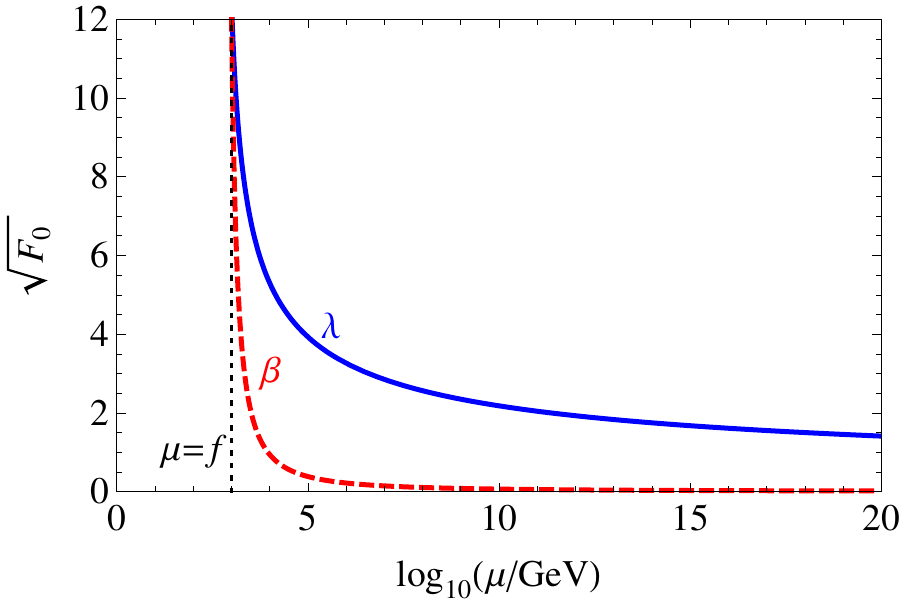}
\includegraphics[width=4.5cm,clip]{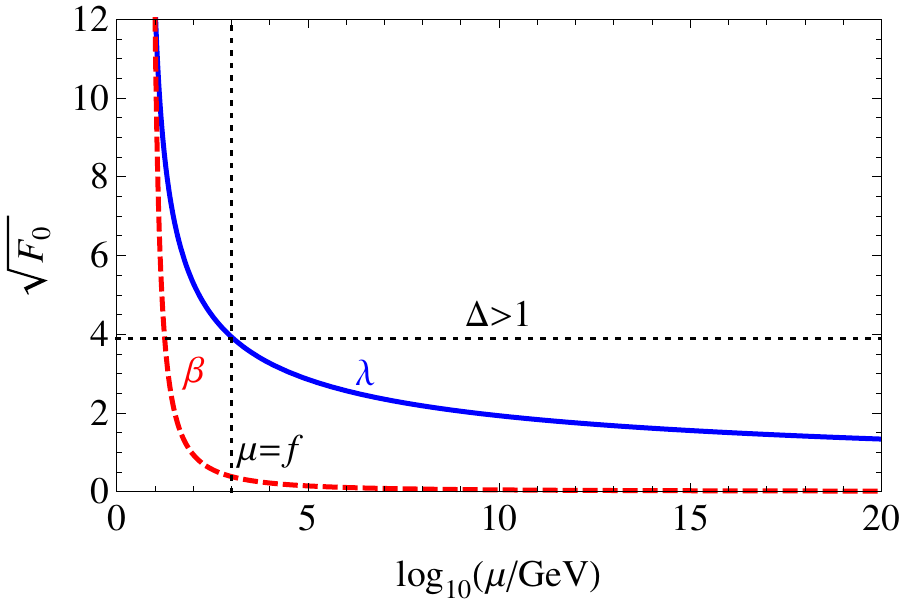}
\includegraphics[width=4.5cm,clip]{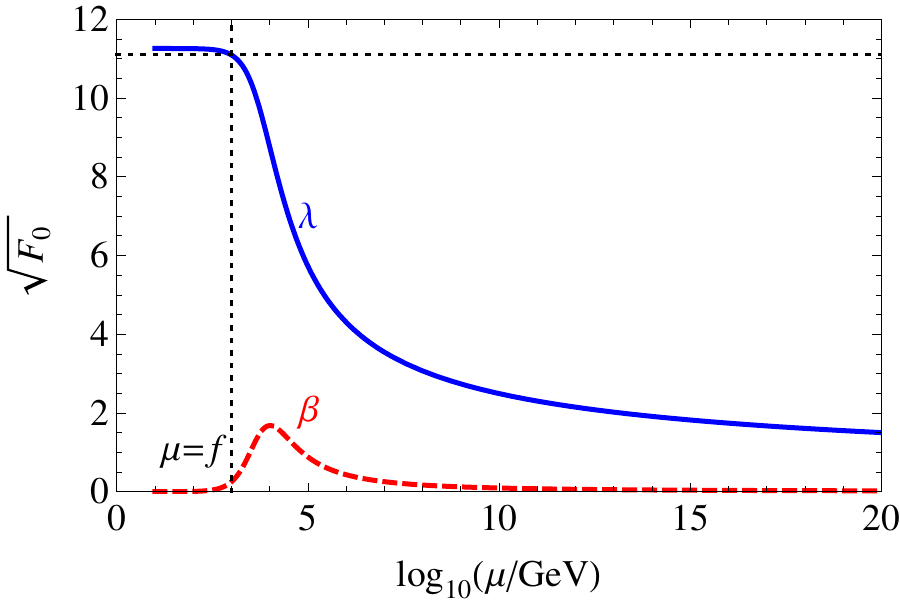}
\caption{Evolution of the perturbation $\lambda$ (solid blue) with the energy scale $\mu$ ($\chi$ in the potential), and of its $\beta$-function (dashed red), for the three different scenarios discussed in the text: 
(up) QCD type $\beta$-function, with the minimum found at $\lambda(f) \! \sim \! \sqrt F_0 \!\sim\! 4 \pi$, 
(middle) the same but for tuned $F_0$ and thus semi-perturbative $\lambda(f)$, and 
(down) Banks-Zaks type $\beta$-function in the non-perturbative regime, allowing $\lambda(f) \! \sim \! \sqrt F_0 \!\sim\! 4 \pi$ while small $\beta(f)$.}
\label{fig-lambda}
\vspace{-0.5cm}
\end{figure}

The first question to address is if the dilaton can naturally weight $\sim \! 125 \, \mathrm{GeV}$.
Being a pseudo-Goldstone boson, one should expect $m_d^2 \ll \Lambda_{IR}^2$, but as outlined before, the very same mechanism leading to SBSI imposes very particular requirements on the dynamics.
All the relevant physics can be extracted from contrasting the minimization condition and the dilaton mass,
\begin{equation}
0 = \left. \frac{dV}{d\chi} \right|_{\chi=f}  = f^3 \left( 4 F+\beta F' \right) \ ,
\label{Vmin}
\end{equation}
\begin{equation}
m_d^2 = \left. \frac{d^2V}{d\chi^2} \right|_{\chi=f} = 4 f^2 \beta F' + O(\beta^2) \ ,
\label{Vmass}
\end{equation}
where all functions on the r.h.s.'s, $F$, $F' \!=\! dF/d\lambda$, and $\beta$, are evaluated at the scale $f$ through their dependence on $\lambda(f)$.
As expected from symmetry arguments, $m_d^2$ is proportional to the explicit breaking, that is the $\beta$-function.
The relevant question is then what is the natural value of $\beta$ at the scale $f$, 
taking into account that the minimization condition should be satisfied for natural values of the parameters.
This crucially relies on the quartic coupling, $F[\lambda] = F_0 + F_1 \lambda + O(\lambda^2)$. 
At face value $F \sim F_0 \sim (4\pi)^2$, thus $\beta \sim 4 \pi$ and no light dilaton is expected.
Scale invariance is badly broken at the condensation scale.
Another option is $F_0 \sim (4\pi)^2/\Delta$, with $\Delta$ a parametrization of the amount of tuning.
The minimization condition Eq.~(\ref{Vmin}) can then be satisfied with perturbative values of $\lambda$, for which $\beta$ and thus $m_d^2/\Lambda_{IR}^2 $ are small.
The minimum is found at $\lambda(f) \simeq -F_0/F_1 \simeq 4\pi/\Delta$.
The amount of fine-tuning required to reproduce $m_d \simeq 125 \, \mathrm{GeV}$ is approximately $\Delta \gtrsim 2 \Lambda_{IR}/m_d \simeq 50$ for $f \!=\! v \!=\! 246$ GeV.
There is however one last natural possibility, that is $\beta/\lambda \ll 1$ even when $\lambda$ becomes non-perturbative. If this non-trivial dynamical property is present, the minimization condition $F \!\simeq\! 0$ will be inevitably satisfied at some low-energy scale, and the dilaton will be light.\footnote{This might involve the same dynamics aimed in walking technicolor theories, originally proposed as a means to alleviate flavor problems \cite{Appelquist:1986an}.}
These three possibilities for the perturbation and its running are sketched in Figure~\ref{fig-lambda}.
The plots should be taken as simple representations of the aforementioned behaviors.
Besides, for the compelling case of $\lambda$ growing strong but keeping a small $\beta(\lambda)$, we show (downmost figure) the instance of a coupling evolving from a UV (trivial) fix point to an IR (strongly coupled) fix point.
This is certainly not the only realization of a parametrically suppressed $\beta$-function, generically $\beta(\lambda) = \epsilon \, b(\lambda)$, with $\epsilon \ll 1$.

It is worthwhile to notice that the AdS/CFT correspondence provides a calculable implementation of these alternatives in the context of a warped extra-dimension \cite{Randall:1999ee}.
While the stabilization mechanism proposed by Goldberger and Wise in \cite{Goldberger:1999uk} tunes $F_0 = 0$, it has been recently showed in \cite{Bellazzini:2013fga} that the more natural realization of SBSI with large $F_0$ is also attainable,
where the perturbation with small $\beta$-function is modeled by a bulk scalar Goldstone boson.

To conclude this section, let us point out that in the SM there are already couplings which contribute to the dilaton mass, the most relevant a priori being the top Yukawa, $y_t$, and gauge interactions, $g',g,g_s$.
A rough estimate for the contribution of the former is
\begin{equation}
( \delta m_d^2 )_{top} \sim - 8 \gamma_t \frac{N_c y_t^2 m_T^2}{16 \pi^2} \simeq
(110 \, \mathrm{GeV})^2 \left( \frac{m_T}{2 \, \mathrm{TeV}} \right)^2 \left( \frac{\gamma_t}{-0.02} \right) \ ,
\label{mdtop}
\end{equation}
where $\gamma_t = \beta(y_t)/y_t$ and $m_T = g_T f$ is the mass of the composite resonances through which the top couples to the strong sector. This is formally a two-loop contribution, and has the right size.

\section{Linear dilaton couplings}
\label{sec-2}

The other piece of information from the LHC regards the linear couplings of the higgs to SM fields.
To compare with experiment and between theory predictions, it is convenient to use the effective lagrangian parametrization
\begin{equation}
\mathcal{L}_h^{(0)} = \frac{h}{v} \left( c_V \, m_V^2 V_\mu V^{\mu}
- c_\psi m_f \overline{\psi} \psi \right) \ ,
\label{Lhp0}
\end{equation}
\begin{equation}
\label{Lhp2}
\mathcal{L}_h^{(2)} = \frac{h}{v} \left(
c_{Z\gamma} Z_{\mu \nu} \gamma^{\mu \nu} +
\frac{c_{\gamma \gamma}}{2} \gamma_{\mu \nu} \gamma^{\mu \nu} +
\frac{c_{gg}}{2} G_{\mu \nu}^a G^{a \mu \nu} 
+ \dots \right) \ ,
\end{equation}
where $V \!=\! W^\pm, Z$, and $\psi \!=\! u,d,l$, and
where we have split the $O(\partial^0)$ couplings, Eq.~(\ref{Lhp0}), from the $O(\partial^2)$ ones, Eq.~(\ref{Lhp2}).
The dots in the latter stand for operators at the same order in derivatives but that contribute to subdominant effects, for instance 3-body $V \psi \overline{\psi}$ higgs decays \cite{Contino:2013kra}.

The couplings of the dilaton are entirely dictated by scale invariance and its breaking \cite{Bellazzini:2012vz}. 
They depend on $\xi \equiv v^2/f^2$, the ratio between the EW scale and the Goldstone decay constant, and on the anomalous dimensions of the SM operators, specifically of the Yukawa coupling for the fermion $\psi$, $\gamma_\psi$, and of the gauge field strength tensors, $\gamma_{g_i} \! = \! (b^{(i)}_{UV}-b^{(i)}_{IR})g_i^2/(4\pi)^2$.
The dilaton predictions for the different $c$-coefficients are shown in Table~\ref{couplings}, along with the SM ones and those of the minimal composite higgs model (MCHM) \cite{Agashe:2004rs}.
\begin{table}[h]
\centering
\caption{Coefficients of the linear higgs couplings in Eqs.~(\ref{Lhp0},\ref{Lhp2}), for the SM, the dilaton, and the MCHM.}
\label{couplings}
\begin{tabular}{cccc}
\hline
coefficient & SM & dilaton & MCHM \\ \hline
$c_V$ & 1 & $\sqrt \xi$ & $\sqrt{1-\xi}$ \\ 
$c_\psi$ & 1 & $(1+\gamma_\psi) \sqrt \xi$ & $\frac{1-(1+n_\psi) \xi}{\sqrt{1-\xi}}$ \footnotemark[3] \\ 
$c_{\gamma \gamma}$ & 0 & $\frac{\alpha}{4 \pi} ( b_{IR}^{(EM)}-b_{UV}^{(EM)} ) \sqrt \xi$ & 0
\vspace{0.1cm} \\ 
$c_{Z \gamma}$ & 0 & 
$\frac{\alpha}{4 \pi t_W} ( b_{IR}^{(2)}-b_{UV}^{(2)} ) \sqrt \xi$ \footnotemark[4]
& 0
\vspace{0.1cm} \\ 
$c_{gg}$ & 0 & $\frac{\alpha_s}{4 \pi} ( b_{IR}^{(3)}-b_{UV}^{(3)} ) \sqrt \xi$ & 0
\vspace{0.05cm} \\ 
\hline
\end{tabular}
\end{table}
\footnotetext[3]{$n_\psi$ comes from the $h$-dependence of the fermion mass, $m_\psi(h) \propto \sin(h/f) \cos^{n_\psi}(h/f)$, with $m_W(h)=gf\sin(h/f)/2$.}
\footnotetext[4]{$O(t_W)$ terms, with $t_W \equiv \tan \theta_W$, have been neglected.}

\noindent All dilaton coefficients carry a model independent $\sqrt \xi$ suppression, 
signal of the fact that the dilaton resembles the SM higgs in the non-decoupling limit $f \to v$, along with small anomalous dimensions $\gamma \ll 1$,
a condition already required to keep the dilaton light.
This is in contrast with composite higgs models, where all deviations from the SM vanish for $\xi = 0$.

\begin{figure}
\centering
\includegraphics[width=6.5cm,clip]{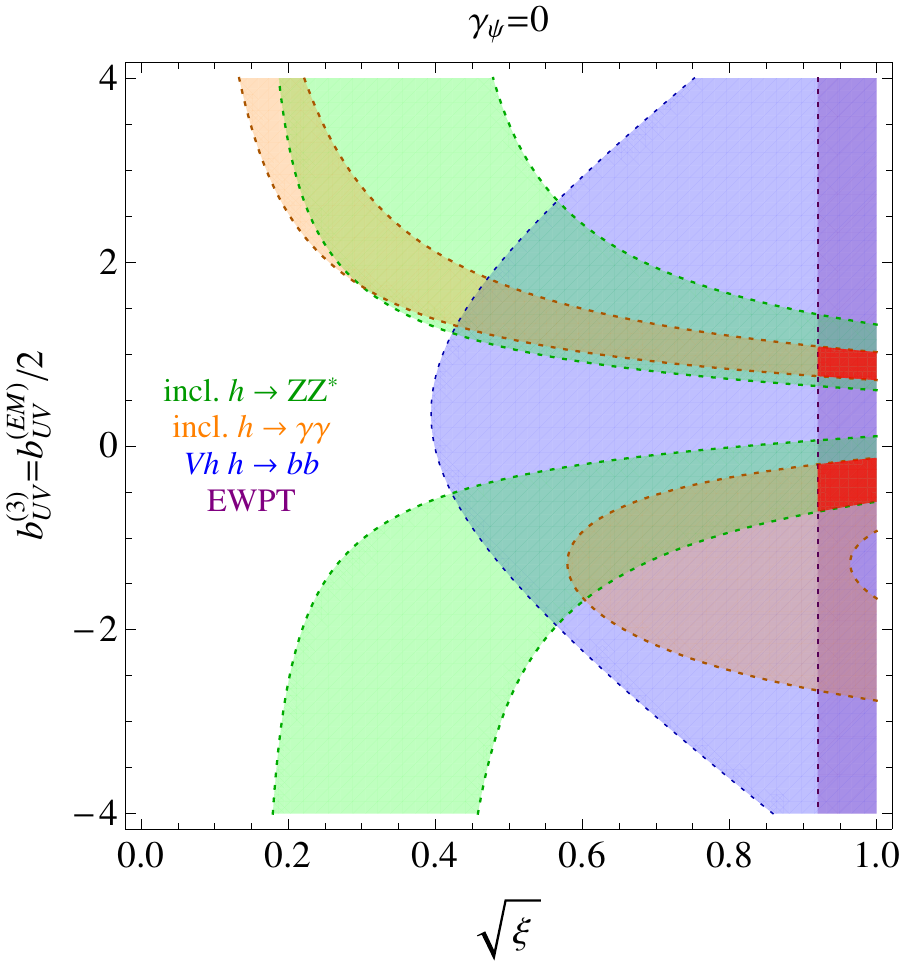}
\caption{Contraints on dilaton parameters, $\sqrt \xi$ and gauge $\beta$-function coefficients $b_{UV}^{(3)} = b_{UV}^{(EM)}/2$, from LHC7 data, at the $2\sigma$ CL, and electroweak precision tests (EWPT), at 99\% CL. 
The IR contribution from composite states to the gauge $\beta$-functions, $b_{IR}^{(i)}$, have been fixed under the assumption that the Goldstone's eaten by the $W$ and $Z$ and the right-handed top are composite.
All Yukawa anomalous dimensions $\gamma_\psi$ have been fixed to zero.
The strongest constrain on $\xi$ comes from EWPT, due to the modified coupling to EW gauge boson, $c_V \neq 1$.
It should be noticed that constraints from LHC8 will be quite stronger, given the consistency of the data with a SM higgs.
See also \cite{Chacko:2012vm}.}
\label{fig-fit}
\end{figure}

The constraints on the dilaton couplings from EWPT and LHC7 data are shown in Figure~\ref{fig-fit}, from where we conclude that $v/f \lesssim 0.9$ and $\gamma \ll 1$ are experimentally favored.
The latter condition is generically satisfied since the anomalous dimensions are formally of one-loop size.
However, the SM higgs effective interactions with gluons and photons arise at one loop as well, thus the dilaton could display $O(1)$ deviations in such couplings.
To avoid it the strong sector should have modest SM central charges.
Finally, the requirement of a small separation between $v$ and $f$ seems to point towards non-trivial  dynamics generating VEV's for EW-charged operators only (among the dimensionful ones), in particular those with the quantum numbers of the SM Higgs doublet field.

We conclude this section by pointing out that the genuine effect of a composite EW scale would be the growth of scattering amplitudes with energy. In this regard, notice that the dilaton accidentally unitarizes $WW$ scattering, the better the closer is $\xi$ to unity, e.g.
\begin{equation}
\label{WW}
\mathcal{A}(W^+ W^- \to ZZ) \simeq \frac{s}{v^2} (1-c_V^2) = \frac{s}{v^2} (1-\xi) \ .
\end{equation}
Given $\xi < 1$, unitarity should ultimately be preserved by the exchange of extra resonances, as in \cite{Bellazzini:2012tv}.

\section{Double dilaton production}
\label{sec-3}

Since the dilaton is able to reproduce the observed higgs-like behavior, the next question is if there are any unavoidable deviations from the SM higgs in other observables,
to be studied at the LHC.
The answer is yes, and the key process is double higgs production.
Using again an effective parametrization of the higgs couplings, the extra relevant interactions are the double higgs couplings to SM fields,
\begin{equation}
\mathcal{L}_{h^2}^{(0)} = \frac{h^2}{v^2} \left( \frac{d_V}{2} \, m_V^2 V_\mu V^{\mu} -
d_\psi m_\psi \overline{\psi} \psi \right) \ ,
\label{Lh2L0}
\end{equation}
\begin{equation}
\mathcal{L}_{h^2}^{(2)} =  \frac{h^2}{v^2} \left(\frac{d_{gg}}{2} G_{\mu \nu}^a G^{a \mu \nu} + \dots \right) \ ,
\label{Lh2L2}
\end{equation}
where the dots stand for terms involving EW gauge bosons, and the trilinear higgs interaction term,
\begin{equation}
\mathcal{L}_{h^3} = - c_3 \frac{1}{6} \left( \frac{3 m_h^2}{v} \right) h^3 \ .
\label{Lh3}
\end{equation}
The corresponding predictions for the SM, the dilaton, and the MCHM are shown in Table~\ref{doublecouplings}.
\begin{table}[h]
\centering
\caption{Coefficients of the double and triple higgs couplings in Eqs.~(\ref{Lh2L0},\ref{Lh2L2}) and (\ref{Lh3}), for the SM, the dilaton, and the MCHM.}
\label{doublecouplings}
\begin{tabular}{cccc}
\hline
coefficient & SM & dilaton & MCHM \\ \hline
$d_V$ & 1 & $\xi$ & $1-2\xi$ \vspace{0.1cm} \\ 
$d_\psi$ & 0 &
$\frac{1}{2} \gamma_\psi \xi$ \footnotemark[5] & $\frac{-\xi(1+3 n_\psi-(1+n_\psi)^2 \xi)}{2(1-\xi)}$ \vspace{0.1cm} \\ 
$d_{gg}$ & 0 & 
$-\frac{\alpha_s}{8 \pi} ( b_{IR}^{(3)}-b_{UV}^{(3)} ) \xi$ \footnotemark[5] & 0 \vspace{0.1cm} \\ 
$c_3$ & 1 & 
$\frac{1}{3} (5+d\beta/d\lambda) \sqrt \xi$ \footnotemark[6] & $\frac{1-(1+\tilde n_\psi) \xi}{\sqrt{1-\xi}}$ \footnotemark[7] \\ 
\hline
\end{tabular}
\end{table}

\noindent Again the dilaton resembles the SM higgs for $\xi \to 1$ and $\gamma \ll 1$, except remarkably in the trilinear interaction, $c_3$.
This can be understood by noticing that the SM result $c_3 \!=\!1$ is reproduced if the perturbation that explicitly breaks the scaling symmetry is a scalar mass term, since then $d\beta/d\lambda = -2$ \cite{Goldberger:2007zk}.
However, the natural realization of the light dilaton hypothesis implies $d\beta/d\lambda \propto m_d^2/\Lambda_{IR}^2$, which makes it a subleading contribution.
This fact appoints double higgs production as a key probe to test the dilaton scenario, with the potential to become the neatest manifestation of the dilatonic nature of the higgs.

In this regard, notice that due to the relation between the linear and double dilaton couplings to EW gauge bosons, the growth with energy in $WW$ scattering to $hh$ is absent at leading order,
\begin{equation}
\label{hh}
\mathcal{A}(W^+W^- \to hh) \simeq \frac{s}{v^2} (d_V-c_V^2) = 0 \ .
\end{equation}
This is an important difference with respect to the composite higgs where, in the high energy regime $\mathcal{A}(W^+W^- \!\to\! hh) \simeq \mathcal{A}(W^+W^- \!\to\! ZZ)$ is expected, due to the higgs being part of an SO(4) vector, contrary to the dilaton. 
Of course the relation Eq.~(\ref{hh}) is affected by higher-order terms in derivatives, for instance $(1 / 16\pi^2 \chi^3) \partial_\mu \chi \partial_\nu \chi \partial^\mu \partial^\nu \chi$ or $2 (m_V^2/\chi^2) V_\mu V_\nu \partial^\mu \chi \partial^\nu \chi$.
Also, notice that the first operator breaks the $h \to - h$ parity symmetry present in the chiral lagrangian of the MCHM. \\

In summary, a composite dilaton is a natural and viable candidate for the higgs-like state discovered at the LHC.
Its linear couplings are expected to deviate from the SM, although not substantially.
Thus double higgs phenomenology sets the path to test the higgs-like dilaton scenario.

\footnotetext[5]{Leading order in anomalous dimensions.}
\footnotetext[6]{Valid for close to marginal perturbations at the scale $f$, $d_{\mathcal{O}_\lambda}-4 \ll 1$.}
\footnotetext[7]{This strongly depends on the higgs potential, taken here to be of the form $V(h) = \cos^{1+\tilde n_\psi}(h/f) \left(\alpha - \beta \cos^{1+\tilde n_\psi}(h/f) \right)$.\vspace{0.2cm}}

\section*{Acknowledgements}

It is a pleasure to thank B.Bellazzini, C.Cs\'aki, J.Hubisz, and J.Terning for the collaboration that led to the main results presented here and published in \cite{Bellazzini:2012vz}.

%

\begin{thebibliography}{}
%
%

\bibitem{Bellazzini:2012vz}
  B.~Bellazzini et al., 
  Eur.\ Phys.\ J.\ C {\bf 73} (2013) 2333
  [arXiv:1209.3299 [hep-ph]].

\bibitem{Agashe:2004rs}
  K.~Agashe, R.~Contino and A.~Pomarol,
  Nucl.\ Phys.\ B {\bf 719} (2005) 165
  [hep-ph/0412089].

\bibitem{Fubini:1976jm}
  S.~Fubini, 
  Nuovo Cim.\ A {\bf 34} (1976) 521.

\bibitem{Coleman:1973jx}
  S.~R.~Coleman and E.~J.~Weinberg,
  Phys.\ Rev.\ D {\bf 7} (1973) 1888.

\bibitem{Appelquist:1986an}
  T.~W.~Appelquist, D.~Karabali and L.~C.~R.~Wijewardhana,
  Phys.\ Rev.\ Lett.\  {\bf 57} (1986) 957.


\bibitem{Randall:1999ee}
  L.~Randall and R.~Sundrum,
  Phys.\ Rev.\ Lett.\  {\bf 83} (1999) 3370,
  hep-ph/9905221.

\bibitem{Goldberger:1999uk}
  W.~D.~Goldberger and M.~B.~Wise,
  Phys.\ Rev.\ Lett.\  {\bf 83} (1999) 4922,
  hep-ph/9907447.

\bibitem{Bellazzini:2013fga}
  B.~Bellazzini, C.~Csaki, J.~Hubisz, J.~Serra and J.~Terning,
  arXiv:1305.3919 [hep-th].

\bibitem{Contino:2013kra}
  R.~Contino, M.~Ghezzi, C.~Grojean, M.~Muhlleitner and M.~Spira,
  arXiv:1303.3876 [hep-ph].

\bibitem{Chacko:2012vm}
  Z.~Chacko, R.~Franceschini and R.~K.~Mishra,
  JHEP {\bf 1304} (2013) 015
  [arXiv:1209.3259 [hep-ph]].

\bibitem{Bellazzini:2012tv}
  B.~Bellazzini et al., 
  JHEP {\bf 1211} (2012) 003
  [arXiv:1205.4032 [hep-ph]].

\bibitem{Goldberger:2007zk}
  W.~D.~Goldberger et al., 
  Phys.\ Rev.\ Lett.\  {\bf 100} (2008) 111802
  [arXiv:0708.1463 [hep-ph]].

\end{thebibliography}
%
%

\end{document}